# Toward Alternative Earths: Habitability of Solar System Bodies at Earth's Orbit

Mohammed Abdel Razek[1]

This paper presents the first structured evaluation of Solar System bodies hypothetically relocated to Earth's orbit (1 AU) to assess their potential as alternative habitats. Using comparative criteria—planetary size and gravity, atmospheric retention, volatile accessibility, weather system potential, soil development feasibility, and orbital transfer cost—we find that most bodies are unsuitable. Mercury and the Moon lack volatiles and atmospheres, while gas and ice giants offer no solid surfaces. Venus, despite strong atmospheric retention, remains constrained by extreme greenhouse forcing. Mars emerges as the most viable candidate, balancing accessibility and volatile resources. Titan provides conditional long-term promise, with a dense atmosphere and rich organics that could transition to a water-based cycle at 1 AU. These findings highlight new pathways for planetary engineering and long-term human survival.

The search for Earth-like exoplanets remains one of the most ambitious goals in modern astronomy. The main idea is based on the region where conditions may allow liquid water to exist. There are two ways to detect exoplanets: the transit method, which measures the tiny dip in a star's brightness when a planet passes in front of it, and the radial velocity method, which detects the gravitational tug of a planet on its host star (Christopher J., 2014). The feasibility of transferring missions to other worlds depends on several interconnected factors. The energy cost, expressed as Δv, includes Earth escape, interplanetary transfer, and orbital capture; Venus and Mars need comparatively low Δv (~5–7 km/s), while Europa and Titan demand far higher values due to their distance and the deep gravity wells of Jupiter and Saturn. Orbital distance from Earth is critical: nearby neighbors such as Venus (~0.28 AU (Earth's orbit)) and Mars (~0.5–2.5 AU) are relatively easy to reach within a few months to a year, whereas distant moons like Europa (5.2 AU) or Titan (9.5 AU) require far longer journeys of 5–10 years and significantly more energy.

Despite these breakthroughs, detecting true Earth analogs remains challenging. Earth-sized planets are small, faint, and often overshadowed by the glare of their host stars. Confirming their atmospheric conditions and surface environments requires technologies still under development. Furthermore, interstellar distances make in-depth exploration impossible with current spacecraft, limiting our knowledge to remote sensing. Nevertheless, the search continues to inspire new strategies and innovations.

The characterization of detected exoplanet atmospheres remains one of the most difficult challenges. Determining whether a planet's atmosphere could support life requires identifying the presence of gases such as water vapor, oxygen, methane, or carbon dioxide. These observations rely on advanced techniques like transit spectroscopy, which analyzes starlight filtered through a planet's atmosphere during a transit, or direct imaging, which attempts to isolate the faint reflected or emitted light of the planet itself. However, these signals are extremely weak for small, Earth-sized planets in the habitable zone and often fall below the sensitivity limits of current telescopes. As Howell notes, even next-generation observatories will face significant hurdles in detecting biosignatures due to the faintness and complexity of planetary atmospheric signals (Howell, 2020).

The definition of "Earth-like" remains central to exoplanetary science. Does "Earth-like" mean a rocky planet located in the habitable zone around its star, orbiting a Sun-like G-type star? Or should it also include features such as plate tectonics, oceans, an oxygen-rich atmosphere, and continents? Some even extend the concept to include human-specific environments, equating "Earth-like" with conditions that support civilization. In this paper, we propose a new definition of "Earth-like" by searching within our solar system for a planet that could be transported into Earth's orbit.

Mars, Venus, Europa, or Titan possess characteristics that could, under the right conditions, support some form of habitability. However, their current orbital positions and environmental conditions are hostile to life as we know it. By hypothesizing the relocation of these worlds into Earth's orbital zone, we open a new framework for thinking about habitability. This framework emphasizes engineering orbital environments rather than searching for naturally occurring Earth analogs far beyond our reach. This approach directly addresses the major barriers of interstellar exploration. Unlike exoplanets located light-years away, the planets and moons of our solar system are accessible with foreseeable space technologies. Studying how these bodies might respond to Earth-like orbital conditions could provide insights into the fundamental requirements for habitability without the need to cross interstellar distances.

[1] Department of Mathematics and Computer Science, Faculty of Science, Al-Azhar University, 18843 Nasr City, Cairo, Egypt. (email: abdelram@azhar.edu.eg).

The primary aim of this paper is to explore the feasibility of creating Earth-like conditions within our own solar system by repositioning certain planets or moons into Earth's orbital zone. Unlike traditional exoplanet research, which focuses on distant and often inaccessible worlds, this study investigates a transformative approach that leverages celestial bodies already within reach.

Specifically, the paper seeks to:

- Model Orbital Transfer Feasibility
  Assess the theoretical dynamics, energy requirements, and engineering challenges associated with relocating a planetary body or moon into Earth's orbit, drawing on orbital mechanics and gravitational modeling.
- Evaluate Environmental Adaptation Potential
  Investigate whether planets such as Mars, Venus, or icy moons like Europa and Titan could, once relocated, adapt to Earth-like orbital conditions—considering factors such as solar radiation, surface temperature, and atmospheric evolution.
- Examine Habitability Prospects
  Explore the potential for relocated planets to develop or sustain life-supporting environments, identifying which candidates within our solar system might be most viable for transformation into habitable worlds.

## Result and discussion

Not every planet or moon within our solar system is a suitable candidate for relocation into Earth's orbital zone. To narrow the scope, clear selection criteria are essential. The evaluation in this study is guided by the following factors:

**Size & Gravity:** Large enough to retain an atmosphere and drive geologic/hydrologic cycles, but not so massive that relocation is infeasible. Fig 1 illustrates why planet size and escape velocity are critical factors for habitability. It plots the radius of planets (relative to Earth) on the horizontal axis and their escape velocity (in km/s) on the vertical axis. Escape velocity is the minimum speed an object needs to break free from a planet's gravitational pull.

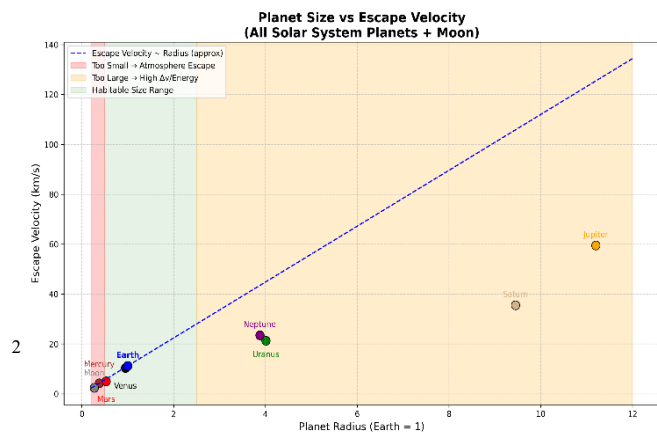

A dashed blue line shows the general trend that larger planets have higher escape velocities, assuming Earth-like density. Each Solar System planet, along with the Moon, is plotted according to its actual measurements. There are three zones highlighted to indicate different habitability conditions. In the red zone, on the left side of the graph, are planets and moons smaller than about half of Earth's radius. These worlds, such as the Moon, Mercury, and Mars, have very low escape velocities and are unable to hold onto thick atmospheres. As a result, they lose gases and water into space, making it extremely difficult to sustain long-term habitability.

In the green zone, between about 0.5 and 2.5 Earth radii, planets are large enough to retain an atmosphere but not so massive that gravity becomes an obstacle for technological access. Earth and Venus fall within this category. Earth represents the ideal balance: it has enough gravity to preserve a breathable atmosphere and liquid water, while still allowing spacecraft to leave the surface with current technology. This size range is often considered the "sweet spot" for habitability.

The orange zone, on the far right of the graph, contains very large planets such as Jupiter, Saturn, Uranus, and Neptune. These planets have extremely high escape velocities, making it impractical for spacecraft to leave their surfaces. In addition, these worlds are mostly gas or ice giants with no solid ground, which further reduces their potential for habitability. However, their moons could still be interesting targets for exploration.

**Atmospheric Retention Potential**

Atmospheric retention potential depends on the balance between gravity (escape velocity), temperature (molecular speed), and volatile supply. A body that is large, cool, and rich in volatiles can maintain or rebuild an atmosphere. A small, hot, or volatile-poor body cannot. The calculation of this paper is under the assumption that any planet sits at 1 AU (Earth's orbit).

To calculate the escape velocity, the Maxwell–Boltzmann distribution. It describes the probability distribution of speeds of particles in a gas at thermal equilibrium. the fraction of Maxwell–Boltzmann molecular speeds exceeding escape velocity (a more direct Jeans-escape style check) for $N_2$ and $CO_2$ at $T = 288$ K for the bodies you asked about.

As shown in the figure 2, Venus, with an escape velocity of 10.36 km/s comparable to Earth's, would have an equilibrium temperature at 1 AU of about 194 K due to its high albedo, although its actual surface temperature is governed by an intense $CO_2$-driven greenhouse effect. Its large volatile reservoir, particularly $CO_2$ and a history of water, indicates strong atmospheric retention potential. At Earth's orbit, a Venus-like planet could sustain a very thick atmosphere, with greenhouse processes still driving high surface temperatures, making its gravity and volatile inventory far more influential than atmospheric losses.

For Venus, the escape-to-thermal speed ratios are about 20.5 for $N_2$ and 25.6 for $CO_2$, indicating clear retention. Mars shows ratios of 9.9 ($N_2$) and 12.4 ($CO_2$), which suggest retention by this simple thermal-speed test, yet its present thin atmosphere reflects other loss processes such as a weak magnetic field, sputtering, and limited volatile supply, meaning this test is necessary but not sufficient. Titan's ratios are around 5.2 for $N_2$ (marginal, allowing slow escape at 1 AU) and 6.5 for $CO_2$ (likely retained), but its current stability relies on its very cold temperature (~94 K); if moved to 1 AU, warming would make $N_2$ more vulnerable. Mercury, with ratios of 8.4 ($N_2$) and 10.5 ($CO_2$), appears capable of retention by thermal criteria, yet cannot maintain an atmosphere due to solar wind stripping, high surface temperatures, lack of volatiles, and no replenishment. The Moon, with ratios of 4.7 ($N_2$) and 5.9 ($CO_2$), is marginal by the test, but practically cannot sustain an atmosphere because of volatile scarcity and constant solar and space weathering.

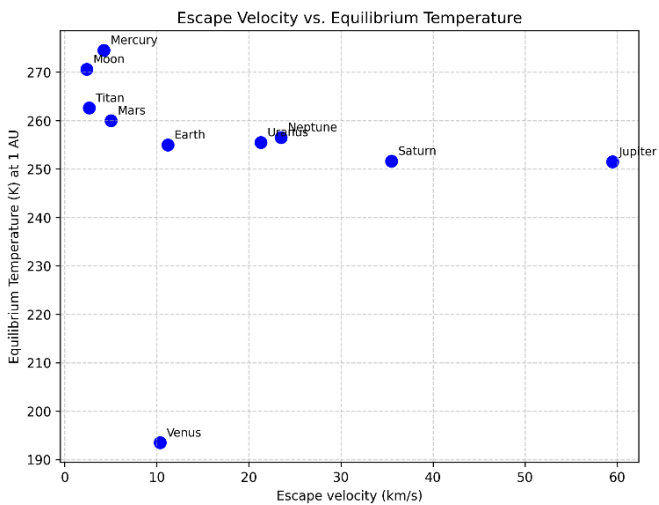

*Figure 3 Escape Velocity vs. Equilibrium*

**Habitability Potential**
Prior evidence of liquid water, subsurface oceans, or complex chemistry (prebiotic organics) that could mature into surface habitability once at 1 AU. Figure 4 shows a comparative table of habitability potential at 1 AU.

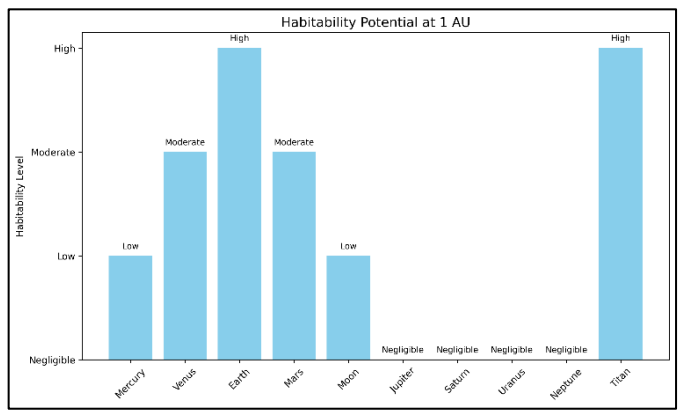

*Figure 4 Habitability potential at 1 AU*

At 1 AU, the potential of Solar System bodies varies widely. Mercury and the Moon rank very low because they lack volatiles, atmospheres, and any history of stable liquid water, leaving them barren even if placed in Earth's orbit. Venus and Mars fall into the moderate category: Venus retains abundant $CO_2$ and $N_2$ and likely had early water, so with reduced greenhouse forcing it could stabilize into a habitable state, while Mars shows strong evidence of past liquid water and subsurface ice, and at 1 AU could sustain habitability more easily with better atmospheric retention. Earth remains the benchmark for habitability, with liquid water, rich volatiles, and a protective atmosphere, making it very high in potential. The gas and ice giants—Jupiter, Saturn, Uranus, and Neptune—offer negligible potential since they lack solid surfaces, though some of their moons could be more interesting at 1 AU. Titan stands out with high conditional potential: it already has a dense atmosphere, abundant organics, and a suspected subsurface ocean, and if moved to Earth's orbit, its methane–ethane cycle could shift to a water-based cycle, opening the door to habitability. In summary, Earth and Titan offer the highest chances, Venus and Mars hold moderate promise, while Mercury, the Moon, and the giant planets are unlikely to support habitable conditions.

**Orbital Distance from Current Location (Transfer Feasibility)**
Closer bodies like Mars and Venus have more feasible chances to think of being transferred with future technology due to lower energy needs and shorter travel compared to the outer moons like Europa and Titan which offer extraordinary habitability potential but come at a steep cost in time, energy, and logistics. The calculation for the distance between planets and the distance to Earth's orbit is taken by taking the absolute difference between each body's semi-major axis and Earth's (1 AU).

Figure 5 shows

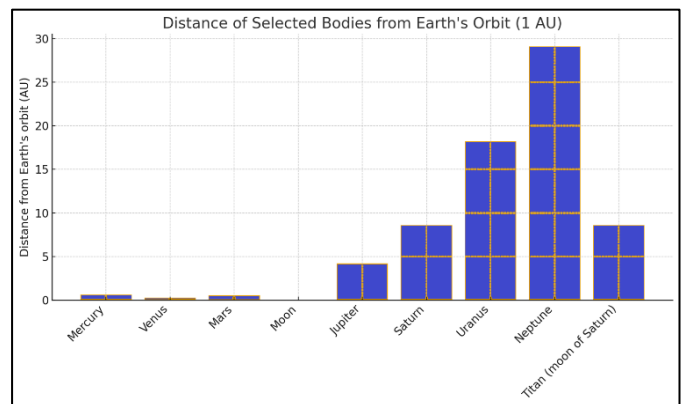

*Figure 5 Distance to Earth's orbit (1 AU)*

Figure 5 compares how far the orbits of different bodies are from Earth's orbit (1 AU). The Moon is essentially at Earth's orbit, only about 0.0026 AU (~384,000 km) away, which explains why it is our most accessible destination. Venus (~0.28 AU, ~41 million km) and Mars (~0.52 AU, ~78 million km) are the closest planets, making them the most feasible targets for near-term interplanetary travel. Mercury, despite being closer to the Sun, is ~0.61 AU (~92 million km) from Earth's orbit, requiring more energy to reach than Venus or Mars. Beyond the inner Solar System, the chart shows a dramatic jump: Jupiter (~4.2 AU, ~629 million km), Saturn and its moon Titan (~8.6 AU, ~1.28 billion km), followed by the distant ice giants, Uranus (~18 AU, ~2.7 billion km) and Neptune (~29 AU, ~4.35 billion km). The steadily increasing bar heights highlight why the Moon, Venus, and Mars are practical destinations, while missions to the outer planets and their moons involve vastly greater challenges in distance, time, and energy.

**Weather System Potential**
under the assumption that each body is relocated to Earth's orbit (1 AU) and left to its own physical properties (same mass, composition, rotation, atmosphere as now) so it receives Earth-like solar flux. Figure 6 shows the estimated chance for a body to have sustained weather system at 1 AU.

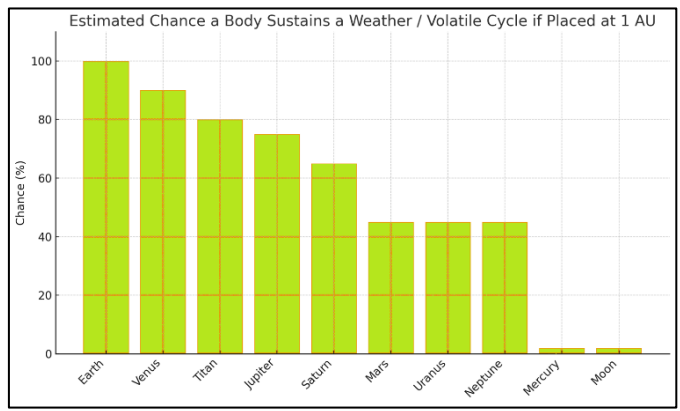

*Figure 6 chance of sustained weather system at 1 AU*

Venus already has a dense atmosphere and strong super-rotation; at 1 AU temperatures would drop dramatically but a massive $CO_2$ atmosphere would retain heat and support vigorous circulation; if some water vapor exists or can be supplied, a strong hydrological cycle (clouds, rain) could develop or be engineered. Mars sits between: it has ice and the potential for a thicker $CO_2$ atmosphere, so it's possible to develop a sustained weather/hydrological system at 1 AU, but it likely needs atmospheric augmentation or release of subsurface volatiles. Mercury is in effect with no atmosphere and very low volatiles; tiny gravity and large temperature swings; would not hold a substantial atmosphere at 1 AU unless atmosphere is engineered or added. Titan already hosts an active volatile cycle (methane) and has a thick atmosphere. Neptune's atmosphere is active due to internal heat; at 1 AU it would be strongly altered but would still sustain powerful circulation and cloud chemistry.

**Potential for Arable Soil Development**

We consider each body orbiting at 1 AU like Earth but keeping its own material composition and chemistry. Now, we try to answer the question *Could we convert local regolith/ice into arable soil (with water, gases, minerals, and living biota like microbes/organics)?* Figure 7 shows Arable soil development potential of selected Solar System bodies if placed at Earth's orbit (1 AU).

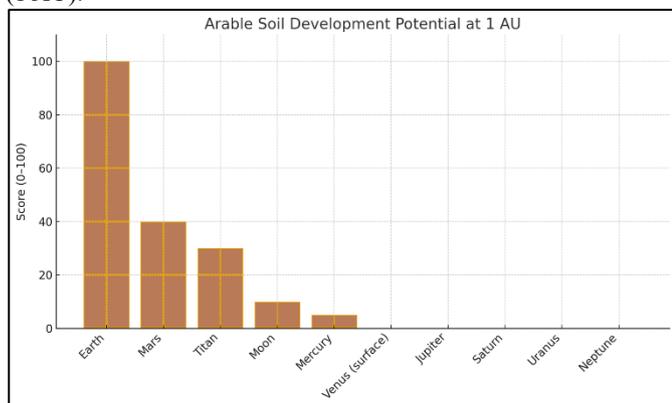

*Figure 7. Arable soil development potential of selected Solar System bodies if placed at Earth's orbit (1 AU).*

Earth serves as the 100% benchmark. Mars and Titan show moderate promise due to available minerals, water ice, or organics, though each requires significant remediation. The Moon and Mercury have minimal potential, while Venus' surface and the gas/ice giants offer virtually no prospects for soil-based agriculture. If each major Solar System body were relocated to Earth's orbit (1 AU), their potential for developing arable soil would vary dramatically. Mercury has very low potential: it lacks a practical atmosphere and almost all water, so agriculture would require massive imports of volatile and organic or fully sealed habitats. Venus (surface) is essentially impossible for soil-based agriculture due to extreme conditions—460 °C surface temperature, crushing pressure, and a corrosive $CO_2$–$SO_2$ atmosphere. Only floating habitats in the cloud layers could be envisioned. Earth, by contrast, already provides the benchmark case with fertile soils, abundant water, living biota, and balanced nutrients. Mars presents a realistically achievable option: its basaltic regolith and accessible water ice can be used to create soil, though challenges include toxic perchlorates, limited organic matter, and low bioavailable nitrogen and phosphorus. These could be overcome with perchlorate removal, ice mining, microbial and organic enrichment, and nutrient supplementation, ideally within domes or greenhouses. The Moon offers only small potential; its dry, abrasive regolith and lack of atmosphere limit prospects, though polar ice pockets may provide water. Successful soil creation would require imported or extracted volatiles, organic inputs, and strong radiation/dust shielding. The gas and ice giants (Jupiter, Saturn, Uranus, Neptune) are not viable, as they have no solid surfaces; any agriculture there would need to occur on floating platforms or, more plausibly, on their moons. Titan, Saturn's largest moon, is conditionally promising: it has abundant organics, a dense nitrogen atmosphere, and an active methane cycle. However, at ~94 K, water is frozen into rock-hard ice. If Titan were warmed at 1 AU and supplied with liquid water, its organics could be processed into humus-like soils, though additional nitrogen and phosphorus inputs would still be necessary.

**Water Availability**

If the major Solar System bodies were relocated to Earth's orbit (1 AU), their potential for volatile resources depends on two key factors: accessibility (how easy it is to reach and use reservoirs of water/ice) and extraction energy (how much effort and power are required to process them). Earth provides the baseline, with abundant accessible water and very low extraction energy. Mars and Titan rank next: both offer significant reservoirs—polar and subsurface ice on Mars, and organics plus water ice on Titan—requiring only moderate energy to process. The Moon and Mercury hold small, localized polar ice deposits that are harder to reach and costly to extract, making them less attractive for large-scale use. Venus, despite its thick atmosphere, provides no accessible water and would demand enormous energy to synthesize it from $CO_2$ and sulfur compounds. The gas and ice giants (Jupiter, Saturn, Uranus, Neptune) are essentially unusable for this purpose, as their volatiles lie buried deep in massive atmospheres under crushing pressures and extreme gravity. In summary, Mars and Titan emerge as the most feasible off-Earth sources of volatiles at 1 AU, while the Moon and Mercury provide limited but useful reserves, and the larger planets are effectively inaccessible.

**Final findings and recommendations**

Based on the above discussion, Mars offers the best balance of feasibility and habitability potential for relocation to Earth's orbit (1 AU). It is the top practical choice now; Titan is the strongest long-term habitability prospect but is much less feasible to relocate. *Relocate Mars to 1 AU* if a goal is to create an additional habitable world within the Solar System that is both scientifically plausible and engineering-feasible in the medium-to-long term. Mars provides the best mix of (i) resource base (water ice and minerals), (ii) moderate transfer cost relative to outer bodies, (iii) reasonable In-Situ Resource Utilization and soil development pathways, and (iv) engineering interventions (artificial magnetospheric shielding, atmosphere augmentation) that are conceptually feasible compared with the near-impossible terraforming required for Venus or the extremely expensive warming of Titan. Use Titan as a long-term objective for habitability (highest conditional biological potential) but only after major advances in propulsion and planetary engineering make the transfer/warming viable.

## Methods

We evaluated major Solar System planets and investigated planets as potential candidates for relocation to Earth's orbital distance (1 AU). The criteria for inclusion required either a solid surface or a known volatile reservoir, ensuring relevance for comparative habitability studies. Major Solar System bodies were selected based on three criteria: (i) ability to gravitationally retain an atmosphere at 1 AU, (ii) presence of volatiles or reservoirs that could be mobilized into a biosphere, and (iii) prior astrobiological interest. This included terrestrial planets (Mercury, Venus, Mars), large moons (Moon, Titan, Europa, Ganymede), and gas/ice giants as contrasting cases. Similar frameworks have been applied in comparative planetology and habitability studies[1].

**Planetary size, gravity, and escape velocity**

Planetary parameters (mass, radius, density) are calculated based on NASA Planetary Fact Sheets[2]. Escape velocity was calculated by

$$v_{esc} = \sqrt{\frac{2GM}{R}} \quad (1)$$

where $G$ is the gravitational constant, $M$ the planetary mass, and $M$ the radius. Classification into "red", "green", and "orange" categories was based on relative Earth radii thresholds, following approaches in exoplanet classification[3].

**Atmospheric retention modeling**

Atmospheric escape was estimated using a Maxwell–Boltzmann framework for thermal escape[4].

$$f(v_{esc}) = 4\pi \left(\frac{m}{2\pi kT}\right)^{3/2} v_{esc}^2 \exp\left(-\frac{mv_{esc}^2}{2kT}\right) \quad (2)$$

where:

$v_{esc}$ is the Escape velocity

$f(v_{esc})$ is the Probability that a molecule has speed $\leq v_{esc}$

$m$ is Mass of one molecule of the gas (e.g., $N_2$ or $CO_2$)

$T$ is the Temperature of the gas

$k$ is the Boltzmann constant ($1.380649 \times 10^{-23}$)

Molecules ($N_2$, $CO_2$) were tested against escape velocities at Earth's mean surface temperature (288 K). Non-thermal escape mechanisms, such as sputtering and ion pickup, were incorporated qualitatively from MAVEN observations of Mars[5].

**Habitability scoring framework**

Each body was scored for prior volatile evidence, geology, and potential for biosphere adaptation at 1 AU. Scores were normalized to Earth's baseline. This follows dynamic habitability assessments in astrobiology[6].

**Orbital relocation feasibility**

We defined a transfer metric as $|a - 1\,\text{AU}|$, where $a$ is the present semi-major axis, to approximate relocation effort. Closer values imply lower energy transfer. This metric is analogous to $\Delta v$–based accessibility analyses used in mission design[7].

**Weather system potential**

Potential for sustaining atmospheric circulation was assessed using atmospheric mass, rotation rate, and volatile phase (liquid water, methane). For gas or ice giants, internal heat flux was also considered. The approach is informed by circulation models of terrestrial and giant planets[8].

**Soil and surface habitability potential**

We assessed regolith composition, hydration potential, and toxicity (e.g., perchlorates on Mars) to evaluate arable soil development. This approach follows studies on in-situ resource utilization (ISRU) for extraterrestrial agriculture[9].

**Volatile reservoir accessibility**

Reservoirs (polar ice, subsurface oceans, methane lakes) were catalogued for each body. Accessibility was qualitatively ranked on a 1–5 scale, accounting for extraction energy and replenishment feasibility. Thermodynamic properties of water ice and hydrocarbons were considered in estimating relative difficulty[10].

**Magnetic protection assessment**

We surveyed intrinsic magnetic fields (Earth, Jupiter, Ganymede) and considered artificial shielding proposals (e.g., dipole fields at L1) for unmagnetized bodies (Mars, Titan, Moon). This follows recent discussions on planetary-scale magnetic shielding for habitability[11].

## Conclusion

This study represents the first structured assessment of in-solar-system planetary relocation as a habitability alternative, systematically evaluating the potential of major Solar System bodies if positioned at Earth's orbit (1 AU). By applying physical, chemical, and environmental criteria—including size and gravity, atmospheric retention, volatile accessibility, weather systems, soil development, magnetospheric shielding, and orbital feasibility—our framework identifies a clear hierarchy of candidates. Earth remains the natural benchmark, while Mars, Venus, and Titan emerge as conditional prospects, each requiring significant planetary engineering to support long-term habitability.

The implications of these findings extend beyond academic speculation: they underscore the importance of planetary-scale strategies for long-term human survival in the face of existential risks. Unlike conventional exoplanet studies, in-system relocation leverages known bodies, predictable resource bases, and proximity, providing a complementary vision for safeguarding human civilization.

Future research should focus on three critical areas: the development of advanced propulsion systems capable of enabling large-scale orbital transfers, in-situ planetary engineering to transform marginal bodies into habitable environments, and the integration of astrobiological and ecological frameworks to ensure the sustainability of transplanted biospheres. Together, these directions offer a roadmap toward expanding humanity's options for survival and flourishing within our own Solar System.

ARTICLE  NATURE COMMUNICATIONS | DOI: 10.1038/ncomms10535
3. Rogers, L. A. Most 1.6 Earth-radius planets are not rocky. Astrophys. J. 801, 41 (2015).
4. Hunten, D. M. Atmospheric escape mechanisms. Planet. Space Sci. 30, 773–783 (1982).
5. Jakosky, B. M. et al. MAVEN observations of the response of Mars to an interplanetary coronal mass ejection. Science 350, aad0210 (2015).
6. Cockell, C. S. et al. Habitability: a review. Astrobiology 16, 89–117 (2016).
7. Chyba, C. F. & Phillips, C. B. Possible ecosystems and the search for life on Europa. Proc. Natl Acad. Sci. USA 99, 2105–2110 (2002).
8. Showman, A. P. et al. Atmospheric circulation of exoplanets and brown dwarfs. In: Exoplanets (ed. Seager, S.) Univ. Arizona Press, 471–516 (2011).
9. Wamelink, G. W. W. et al. Can plants grow on Mars and the Moon: a growth experiment on Mars and Moon soil simulants. PLoS ONE 9, e103138 (2014).
10. Petrenko, V. F. & Whitworth, R. W. Physics of Ice. (Oxford Univ. Press, 1999).
11. Green, J. L. et al. A future Mars environment protection strategy: the artificial magnetosphere approach. Acta Astronaut. 146, 290–296 (2018).
6  NATURE COMMUNICATIONS | 7:10535 | DOI: 6.1038/ncomms10535 | www.nature.com/naturecommunications